\documentclass[conference]{IEEEtran}
\usepackage{cite,subfigure}
\usepackage{amsmath,amssymb,amsfonts,amsthm}
\usepackage{textcomp}
\usepackage{xcolor}
\usepackage{booktabs} 
\usepackage{makecell} 
\usepackage{pgf,pgfkeys,tikz,circuitikz,float,bm,caption}
\usepackage[ruled,vlined]{algorithm2e}
\usepackage{graphicx,multirow}%
\usepackage[super]{nth}
\usepackage{steinmetz,lipsum}
\usepackage{siunitx,upgreek}
\def\BibTeX{{\rm B\kern-.05em{\sc i\kern-.025em b}\kern-.08em
	T\kern-.1667em\lower.7ex\hbox{E}\kern-.125emX}}

\DeclareSIUnit{\rad}{rad}

\graphicspath{{figures/}}

\begin{document}

\title{Waveform Manipulation Against DNN-based Modulation Classification Attacks}
	
\author{ Dimitrios Varkatzas and Antonios Argyriou\\ Department of Electrical and Computer Engineering, University of Thessaly, Greece\\
Email: dvarkatzas@uth.gr, anargyr@uth.gr}

\markboth{Submitted to~~~}{Argyriou \MakeLowercase{\textit{et al.}}: Privacy Threats from Passive Wireless Radar}

\maketitle

\begin{abstract}				
In this paper we propose a method for defending against an eavesdropper that uses a Deep Neural Network (DNN) for learning the modulation of wireless communication signals. Our method is based on manipulating the emitted waveform with the aid of a continuous time frequency-modulated (FM) obfuscating signal that is mixed with the modulated data. The resulting waveform allows a legitimate receiver (LRx) to demodulate the data but it increases the test error of a pre-trained or adversarially-trained DNN classifier at the eavesdropper. The scheme works for analog modulation and digital single carrier and multi carrier orthogonal frequency division multiplexing (OFDM) waveforms, while it can implemented in frame-based wireless protocols. The results indicate that careful selection of the parameters of the obfuscating waveform can drop classification performance at the eavesdropper to less than 10\% in AWGN and fading channels with no performance loss at the LRx.
\end{abstract}

	
\section{Introduction}
One challenging problem in wireless communication systems is learning the digital or analog modulation used in an unknown signal. The term used for this problem is Modulation Classification (MC). MC is a problem that has been studied extensively in the past few years~\cite{Liu18,Oshea18,Zhou19,Wang20,pappas2020,Park21}, and finds applications in the civilian and military domains. Knowing the modulation type of a wireless communication signal is invaluable for military applications since it can allow fingerprinting of the source. For civilian applications MC serves a bigger objective that is concerned with learning the precise characteristics of spectrum utilization, compliance, etc.

At the same time Deep Learning (DL) has taken by storm several practical problems in signal processing, wireless communications and networks. For example DL has been applied successfully in image classification and natural language processing. The success of DL algorithms is tied to a class of applications for which models for driving algorithms do not exist, but instead we have access to large data sets. One of these applications is MC which is very challenging to be addressed with classic model-based signal detection techniques. As a result, DL has been considered for MC through a variety of different Deep Neural Networks (DNNs)~\cite{Oshea18}. Extensive research has been done on the performance of MC with DL for different types of channels that include fading and receiver processing~\cite{Wang20}. These systems either use baseband In-phase and Quadrature-phase (I/Q) complex samples, or generate spectrograms from them and feed them to the DNN classifier. All these systems have shown to offer great accuracy. 

\begin{figure}[t]
	\centering
	\includegraphics[width=0.99\linewidth]{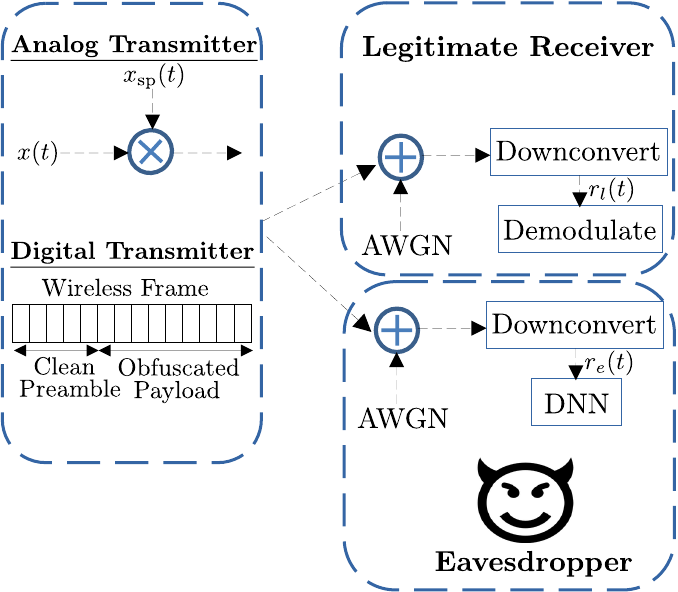}
	\caption{In our scenario a wireless eavesdropper tries to classify a received communication signal $r_e(t)$ with a DNN. When analog communication is used the transmitter mixes the baseband signal $x(t)$ with an obfuscating waveform before upconversion and transmission. For digital transmission only the complex baseband symbols of the payload are mixed with the obfuscating waveform $x_\text{sp}(t)$. }
	\label{fig:topology-for-amc-spoofing}
\end{figure}

However, this effectiveness of DNNs for MC is not always desirable. The scenario that is of interest in this paper, and illustrated in Fig.~\ref{fig:topology-for-amc-spoofing}, consists of a nominal communication pair between a transmitter (Tx) and a legitimate receiver (LRx). The attacker/adversary is a passive eavesdropper that executes the DNN-based MC algorithm. From the perspective of the Tx, this eavesdropper should not be able to learn the used modulation in the payload of the transmitted digital wireless frame or in the analog modulated waveform. However, the LRx should be able to demodulate the received signal without any performance loss. Hence, we desire to defend against an adversary that is an eavesdropper and uses DNN-based MC. 

Recent work on adversarial DL focuses on adapting the learning process when malicious input is given  to the deep network~\cite{papernot16}. DNN-based MC has also been investigated in the context of adversarial learning. Most the works focus on deriving the necessary perturbation to yield a lower classification performance by considering that the adversary is the injector of the perturbation~\cite{Usama19,Bahramali21,Wang23,Liu23}. Then, adversarial training of the DNN is responsible for training under different perturbations so that the classifier is ready to face the threat, i.e. the test data. However, unlike the aforementioned related work our threat model we described in Fig.~\ref{fig:topology-for-amc-spoofing} is quite different: We consider that \textit{the injector of the perturbation is the friendly transmitter and the adversary is the DNN algorithm at the eavesdropper}. Under this different setting a few recent works have focused on hiding the modulation, a method called \textit{modulation obfuscation}. In~\cite{Xiong15,Rahbari23} the authors presented methods to obfuscate quadrature amplitude modulated (QAM) symbols at the transmitter. This is accomplished by embedding the symbols from a lower order to a higher order QAM constellation, hence they obfuscate the modulation order and not the type. So these methods produce QAM symbols that can still be distinguished from other modulations, digital (e.g. frequency shift keying) or analog (FM, AM). In another recent work the authors perturbed digitally modulated symbols at the encoder so that they fool the DNN, but they were not concerned with analog waveforms and OFDM or the precise signal recovery at a LRx~\cite{Gunduz21}. 

\textit{Consequently, the basic problem we set to address is to find the type of perturbation that we should introduce so that the LRx does not experience performance loss for any type of modulation (digital or analog), but the eavesdropper experiences low classification performance.} To solve this problem we set the following requirements for our system: 1) For digital modulation the scheme should be implementable in frame-based wireless protocols, 2) the perturbing signal should be easily and completely removable from the manipulated waveform that the LRx receives, 3) to minimize as much as possible the classification performance at the eavesdropper.

The previous problem is addressed by meeting the defined requirements in ways that we will explain in the rest of this paper. Our contributions with respect to the related work are: 
  \begin{itemize}
 	\item A waveform manipulation strategy that can be implemented as part of both digital frame-based or analog wireless communication and constitutes a defense against DNN-based MC attacks at eavesdroppers.
 	\item A waveform design that is robust to bit decoding at a legitimate receiver, i.e. it is not a randomized un-recoverable perturbation. 
 	\item We present and open-source a new dataset~\cite{6szh-qd43-23} that contains manipulated waveforms and can be used by the community for evaluating newer DNN-based classification attacks against our manipulated waveform.
 \end{itemize}

\section{System Model and Obfuscation Strategy} 
As discussed before, our basic system model considers a communicating pair between a transmitter and legitimate receiver in the presence of an eavesdropper. The eavesdropper is assumed to operate in a way that is consistent with its typical expected capabilities, i.e. it is not part of the communication network and it cannot decrypt the Tx-LRx communication. However, it is reasonable to assume that it knows the carrier frequency $f_c$ so that it downconverts any passband signal (Fig.~\ref{fig:topology-for-amc-spoofing}). This information is trivial to acquire, since it only requires knowledge of the PHY protocol in use (e.g. WiFi at \SI{5}{\giga\hertz}, etc). Hence, MC that we discuss later will take place from baseband samples, an assumption consistent with the literature we discussed. 

\subsection{Meeting the System Design Requirements}
When the information is digitally modulated, the first requirement is satisfied as follows: We insert an obfuscating signal $x_\text{sp}(t)$ only in the payload of a transmitted wireless frame (Fig.~\ref{fig:topology-for-amc-spoofing}) so that when the eavesdropper receives the frame it cannot estimate $x_\text{sp}(t)$ from the preamble (e.g. by treating it as a channel impairment). For continuous-wave analog modulations the eavesdropper could not possibly use known preambles for estimating the perturbation. So $x_\text{sp}(t)$ is mixed with the complete analog baseband waveform $x(t)$ (Fig.~\ref{fig:topology-for-amc-spoofing}).

Regarding the second requirement, i.e. easy and complete perturbation removal at the LRx, it is satisfied by selecting a class of deterministic obfuscating signals that do not include any amplitude variations (e.g. act as a wireless fade) but only time-dependent phase variations. This class of signals was proposed in~\cite{jnl_2023_access} with the purpose of smearing the resulting spectrogram of a wireless digital communication signal.\footnote{That work was not concerned with modulation obfuscation and the performance of the communication subsystem.} Regarding the LRx we explore two cases, one where it does not know $x_\text{sp}(t)$ (and so it experiences some demodulation performance loss) and one case where it knows its parameters and can so it can re-create $x_\text{sp}(t)$ and easily equalize it and remove its effect. The parameters of the obfuscating signal $x_\text{sp}(t)$ are encrypted and transmitted to the LRx.

The third requirement is addressed in the rest of the paper and is achieved by selecting the parameters of the obfuscating signal in a specific way. 
		
\subsection{Waveform Manipulation}
Now we describe how we manipulate the waveform and produce the datasets that are used at the eavesdropper and the LRx. Let $x(t)$ be the baseband modulated signal. We want to introduce a continuous-time perturbation in the transmitted signal so that the eavesdropper missclassifies $x(t)$. Formally, at the Tx we want to solve the following optimization problem for a time $t$: Find the perturbation $p(t)$ that maximizes the loss (classification error) over the training dataset $\mathcal{D}$:
\begin{align}
\max_{p(t)} \sum_{x(t)\in \mathcal{D}}\mathcal{L}\Big ( f \big ( x(t)+p(t) \big ),f ( x(t) ) \Big ) ~~\text{s.t.}~~\|p(t)\|_q\leq \epsilon.
\label{eqn:adversarial-training-1}
\end{align}	
In the above $\mathcal{L}$ is the loss function used for the non-linear DNN classifier $f()$, $\|\cdot\|_q$ is the $L_q$ norm and $\epsilon$ is the perturbation budget. Instead of solving this problem for deriving the perturbation $p(t)$ (e.g. with gradient descent variants~\cite{Madry19}) we propose a specific deterministic waveform that we will show that it offers excellent performance. In particular we select the perturbation in the time domain to be $p(t)=(x_\text{sp}(t)-1)x(t)$, where we name $x_\text{sp}(t)$ the obfuscating waveform. With this configuration the data that will be given as input to the classifier (without any channel impairment yet) is $x(t)+p(t)=x(t)x_\text{sp}(t)$. Note that the last expression means that it can be implemented by mixing the two waveforms (Fig.~\ref{fig:topology-for-amc-spoofing}). Regarding the perturbation budget $\epsilon$, related work on adversarial DNN training tries to maintain its value as small as possible so that the perturbation is undetectable. However, here there is no such requirement since as a Tx we want our modulated data to be miss-classified and we do not care if the eavesdropper detects our perturbation. The only requirement we have is that the power of the transmitted signal is not affected which has the dual benefit of no power waste plus no impact on the signal-to-noise ratio (SNR): The instantaneous power of the transmitted signal should be $|x(t)x_\text{sp}(t)|^2=|x(t)|^2$.

\textbf{Obfuscating Waveform:} An oscillating sinewave that was proposed in~\cite{jnl_2023_access} is selected so that it distorts the spectrograms of the generated signals. This signal denoted as $x_\text{sp}(t)$ is defined to be an frequency-modulated (FM) waveform with maximum instantaneous frequency shift $\delta f$, and frequency $f_m$: 
\begin{align}
	x_\text{sp}(t)=e^{j\frac{\delta f}{f_m}\sin(2\pi f_m t)}.
		\label{eqn:spoofing-signal}
\end{align}
The instantaneous frequency is $f_i(t)=f_c + \delta f \cos(2\pi f_m t)$. As a result, the signal generated by this waveform oscillates in the frequency domain between the maximum instantaneous frequencies -$\delta f$ and +$\delta f$ at a rate of $f_m$~\SI{}{\hertz}. The result of waveform manipulation in this way is that $x(t)$ is experiencing a time-varying phase shift, but the power of $x(t)x_\text{sp}(t)$ is the same as $x(t)$. These two parameters of the obfuscating signal are encrypted and transmitted out-of-band to the LRx.

\textbf{Channel Model:} Since a practical channel is not perfect, the LRx and the eavesdropper will receive a distorted form of the manipulated waveform. To derive the actual signal model, let $h$ be a complex constant that characterizes path loss and any flat fading channel gain, and $n(t)$ be an additive white Gaussian noise (AWGN) process reflecting thermal noise. Then the baseband received signal is~\cite{jnl_2023_access}:
\begin{align}
r(t) = hx_\text{sp}(t) x(t) + n(t).
\label{eqn:signal-model-1}
\end{align}
Note that we may use the notation $r_e(t)$ and $r_l(t)$ to distinguish the received signal at the eavesdropper and the legitimate receiver respectively, when necessary. When the impulse response of the channel is not flat, it is equal to $h(\tau)$ so in this case the output signal model is:
\begin{align}
r(t) = \int_{0}^{t} h(\tau)x_{sp}(t-\tau) x(t - \tau)d\tau + n(t).
\end{align}
	
\textbf{OFDM Model:} While the first dataset we produced used single-carrier modulation, the signals for the second dataset are QAM symbols that were modulated with OFDM. With $N$ subcarriers that are spaced relative to the carrier $f_c$ at locations $f_k = k\Delta f$~\SI{}{\hertz} that can contain data, pilot symbols, or a combination of both, the desired baseband OFDM symbol in continuous time is~\cite{jnl_2023_access}:
\begin{align}
x(t) = \frac{1}{\sqrt{N}} \sum_{k = 0}^{N - 1} X[k]e^{j2\pi k\Delta ft} , \; 0 \leq t \leq T_N
\end{align}
\noindent $X[k]$ is the complex QAM symbol modulated onto subcarrier $k$, and $T_N = N\Delta f$ is the OFDM symbol duration.

\begin{figure}[t]
	\centering
	\includegraphics[width=0.95\linewidth]{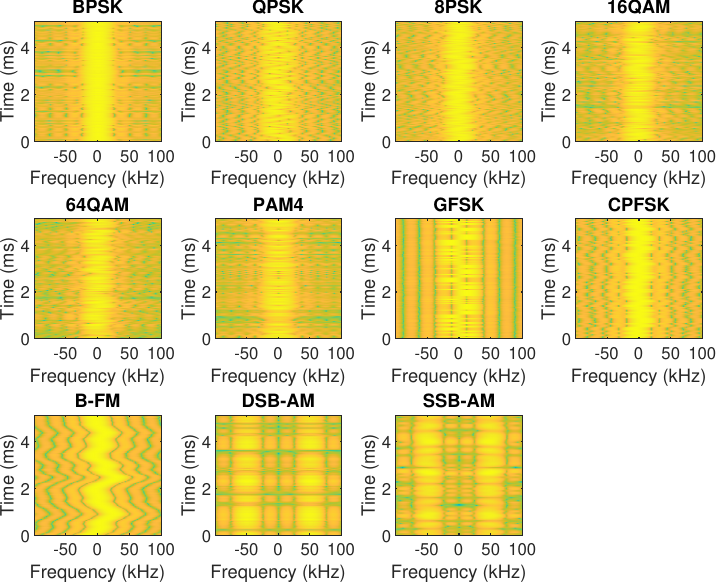}	
	\caption{Spectrograms of the baseline dataset.}
	\label{fig:spectrograms-amplitude}
\end{figure}

\begin{figure}[t]
	\centering
	\includegraphics[width=0.95\linewidth]{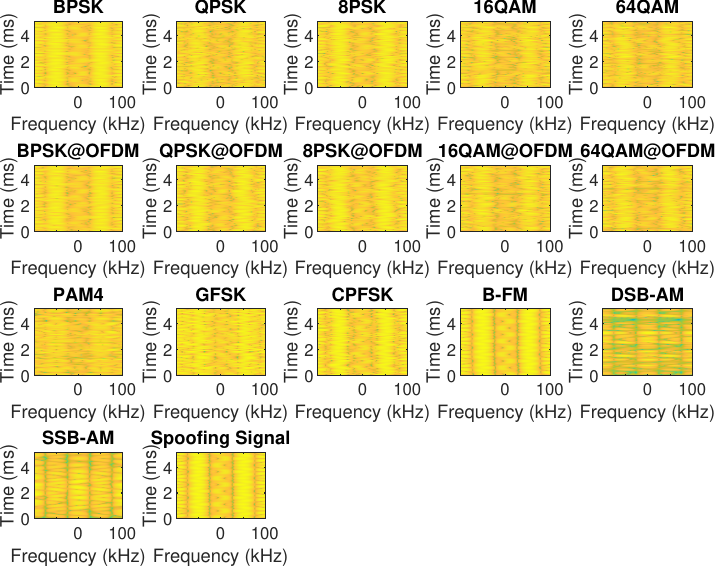}
	\caption{Spectrograms of the received manipulated waveform at SNR=30dB, $\delta f$ = \SI{100}{\hertz}, $f_m$ = \SI{100}{\hertz}.}
	\label{fig:attemp10_Spectrograms}
\end{figure}

\textbf{Perturbation Removal at the LRx:} When the LRx knows the two parameters of the perturbation it can re-create locally $x_{sp}(t)$. Then it can remove it from the received signal in~\eqref{eqn:signal-model-1} with equalization:
\begin{align}
r_{l}(t) x^*_{sp}(t)=hx(t)+n(t)x^*_{sp}(t)
\end{align}
But the AWGN $n(t)$ still remains Gaussian and has the same power since it is simply rotated with the mutiplication with $x^*_{sp}(t)$, i.e. there is no SNR loss.

\section{Baseline and Obfuscated Datasets} 

\textbf{Baseline Dataset for Training:} One existing dataset is used for trainning in this paper which is RadioML2016.10a~\cite{Oshea18}. It contains 11 different types of modulations (eight digital and three analog) in total. This is a very comprehensive dataset that can be used for training a DNN under different conditions. The included single-carrier modulations are B-FM, DSB-AM, and SSB-AM for analog modulation, as well as BPSK, QPSK, 8PSK, QAM16, QAM64, BFSK, CPFSK, and PAM4 for digital modulations. With a normalized average transmit power of 0 dB, data is modulated at a rate of 8 samples per symbol. The full dataset was created using GNU radio as a 1024-sample complex time-domain vector. Through 1024-sample rectangular windowing processing, samples for both datasets are separated into training, validation, and testing. The training examples - each consisting of 1024 samples - are fed into the neural network in 2$\times$1024 vectors with real and imaginary parts separated in complex time samples. 

\textbf{Obfuscated Dataset:} In our case we created a synthetic dataset that contains manipulated waveforms with the communication system model we described in the last section. Radio communication signals are created artificially, and we do it in a way that is identical to a real system by adding modulation, pulse shaping (a raised cosine filter with roll-off 0.5), and different channel models. Besides obfuscation, we add AWGN and time-varying multipath fading for the channel impulse response for a subset of the experiments. Channel bandwidth is \SI{20}{\mega \hertz}. Our dataset does not include additional hardware imperfections since they have been explored in the literature we reviewed. The dataset was created with tools from the Matlab\textsuperscript{\tiny\textregistered} Communications Toolbox. The entire dataset consists of roughly 1.25 GB worth of frames, which are files holding floating point samples~\cite{6szh-qd43-23}. The second part of the dataset is a variation of the first one, in which only QAM symbols are used and are also modulated with OFDM. 

\textbf{Spectral Characteristics:} To illustrate the spectral characteristics of the datasets we present the spectrograms in Fig.~\ref{fig:spectrograms-amplitude} and Fig.~\ref{fig:attemp10_Spectrograms} for the baseline and obfuscated datasets respectively. These graphs show that we can visually distinguish a number of similarities and differences between modulations by examining a single snapshot in the frequency domain. Different obfuscating signal parameters will result in different patterns.

\section{Deep Learning for MC at the Eavesdropper}

\subsection{Deep Neural Network}
Our main method for learning is a Deep Convolutional Neural Network (CNN), which receives a windowed input of the unprocessed baseband time series $r_e(t)$. We employ the sampled version of $r_e(t)$ as a set of $2\times N$ vectors into a narrow 2D CNN. The $N$ I/Q samples make up this 2-wide dimension. State-of-the-art works have shown that a 7-layer CNN maximizes performance~\cite{Oshea18,Liu18} and so we adopt the same architecture.  This CNN consists of six convolution layers and one fully connected layer. Each convolution layer except the last is followed by a batch normalization layer, a rectified linear unit (ReLU) activation layer, and a max pooling layer. In the last convolution layer, the max pooling layer is replaced with an average pooling layer. The output layer has softmax activation. This CNN offers excellent performance for analog and digital modulation as we will soon see.
	
\begin{flushleft}
	\begin{table}[t] 
		\caption{Parameters for Dataset 1.} 
		\label{table:vardescription1}
		\begin{center}
			\begin{tabular}{|*{2}{p{40mm}|}}
				\hline \hline
				
				Modulations of dataset 1 &  BPSK, QPSK, 8PSK, 16QAM, 64QAM, PAM4, GFSK, B-FM, DSB-AM, SSB-AM  \\\hline 
				
				Samples per symbol & 8 \\ \hline 
				Samples per frame & 1024 \\ \hline 
				SNR range & -10 to \SI{30}{\decibel}\\ \hline 
				Number of training samples & 128000 \\ \hline 
				Number of validation samples & 16000 \\ \hline 
				Number of test samples & 16000 \\ \hline 
				Sample rate & \SI{40}{\mega\hertz} \\ \hline 
				
			\end{tabular}
		\end{center}
		
	\end{table}
\end{flushleft}

\begin{flushleft}
	\begin{table}[t]  
		\caption{Parameters for Dataset 2.} 
		\label{table:vardescription2}
		\begin{center}
			\begin{tabular}{|*{2}{p{40mm}|}}
				\hline \hline
				
				Modulations of dataset 2 (OFDM)& BPSK, QPSK, 8PSK, 16QAM, 64QAM  \\\hline 
				
				Samples per symbol & 8 \\ \hline 
				Samples per frame & 1024 \\ \hline 
				SNR range & -10 to \SI{30}{\decibel}\\ \hline 
				Number of training samples & 40000 \\ \hline 
				Number of validation samples & 5000 \\ \hline 
				Number of test samples & 5000 \\ \hline 
				Sample rate & \SI{40}{\mega\hertz} \\ \hline 
				Number of OFDM subcarriers & 64 \\ \hline 
				OFDM cyclic prefix length & 16 samples \\ \hline  
				
			\end{tabular}
		\end{center}
		
	\end{table}
\end{flushleft}	

Stochastic Gradient Decent (SGD) with a mini-batch size of 256 was used as the solver for training the CNN. After several attempts to find the ideal learning rate for the model, we reached a value of approximately 0.02. But nearly after 10 epochs the learning rate of our model needed to be reduced because it was quite high for that epoch. So a learning rate schedule was added according to which after the 9th epoch, the learning rate is reduced by one tenth. It should be stressed that the learning rate interacts with other hyperparameters, such as the mini-batch size and the number of epochs. In this experiment, this interaction may make it hard to isolate the effect of batch size alone on model quality. However, the choice of the mini-batch size influences the training time until convergence, the training time per epoch and the model quality. We determined the appropriate mini-batch to be 256. 
	
Each time series signal was divided into a training, validation, and test sets using a rectangle windowing process with 1024 samples. 80\% of the datasets is used for training, 10\% for validation and the remaining 10\% for testing. There is no general rule of thumb in choosing percentages, because it often depends on the SNR of the data and the size of the training data. However, it is generally true that the more frames we get, the higher the accuracy of the classifier will be. 

\subsection{Training on an Adversarial Dataset}
We also briefly explored a more sophisticated eavesdropper that knows the attack model in~\eqref{eqn:spoofing-signal} but not the individual parameters of this model (that have been sent encrypted to LRx). This model is trained by following exactly the optimization in~\eqref{eqn:adversarial-training-1} but with a minimization objective: The eavesdropper tries to to find the DNN $\hat{f}()$ that minimizes the loss for a dataset that includes perturbations ($x(t)+p(t)$).
	
\section{Evaluation}
We first consider an AWGN channel and explore equalization of the obfuscating signal at the LRx, and then present results for fading channels. All the system parameters for the two datasets we produced are shown in Table~\ref{table:vardescription1} for single carrier modulation (Dataset 1), and Table~\ref{table:vardescription2} for OFDM modulation (Dataset 2). We analyze the DNN performance at the eavesdropper for various pairs of the parameters $\delta f$ and $f_m$ selected for $x_\text{sp}$ by the Tx. An important detail is that these values should not exceed \SI{100}{\hertz}, since the spectra of the resulting waveform must occupy nearly the same bandwidth with $x(t)$. This is because a receiver will filter the frequency band defined by the used standard. Finally, we must note that results are averages and not all modulation types are impacted equally as the SNR and the obfuscating parameters change. 
	
\begin{figure}[t]
	\centering
	\includegraphics[width=0.95\linewidth]{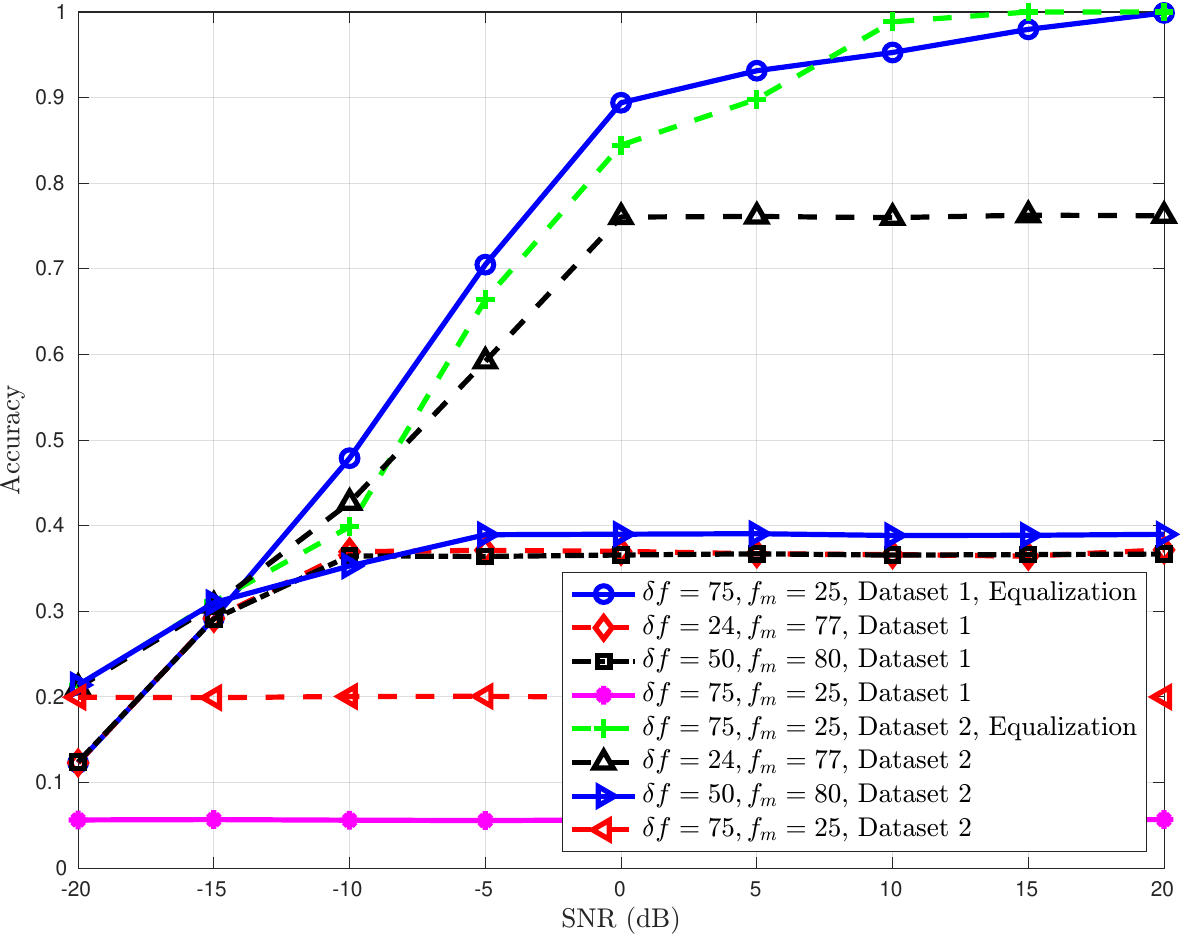}
	\caption{Average classification accuracy in an AWGN channel for all modulations and different parameters in the obfuscating signal. Results with equalization refer to the LRx.}
	\label{fig:accuracy-vs-snr_spoofing}
\end{figure}
\begin{figure}[t]
	\centering
	\includegraphics[width=0.95\linewidth]{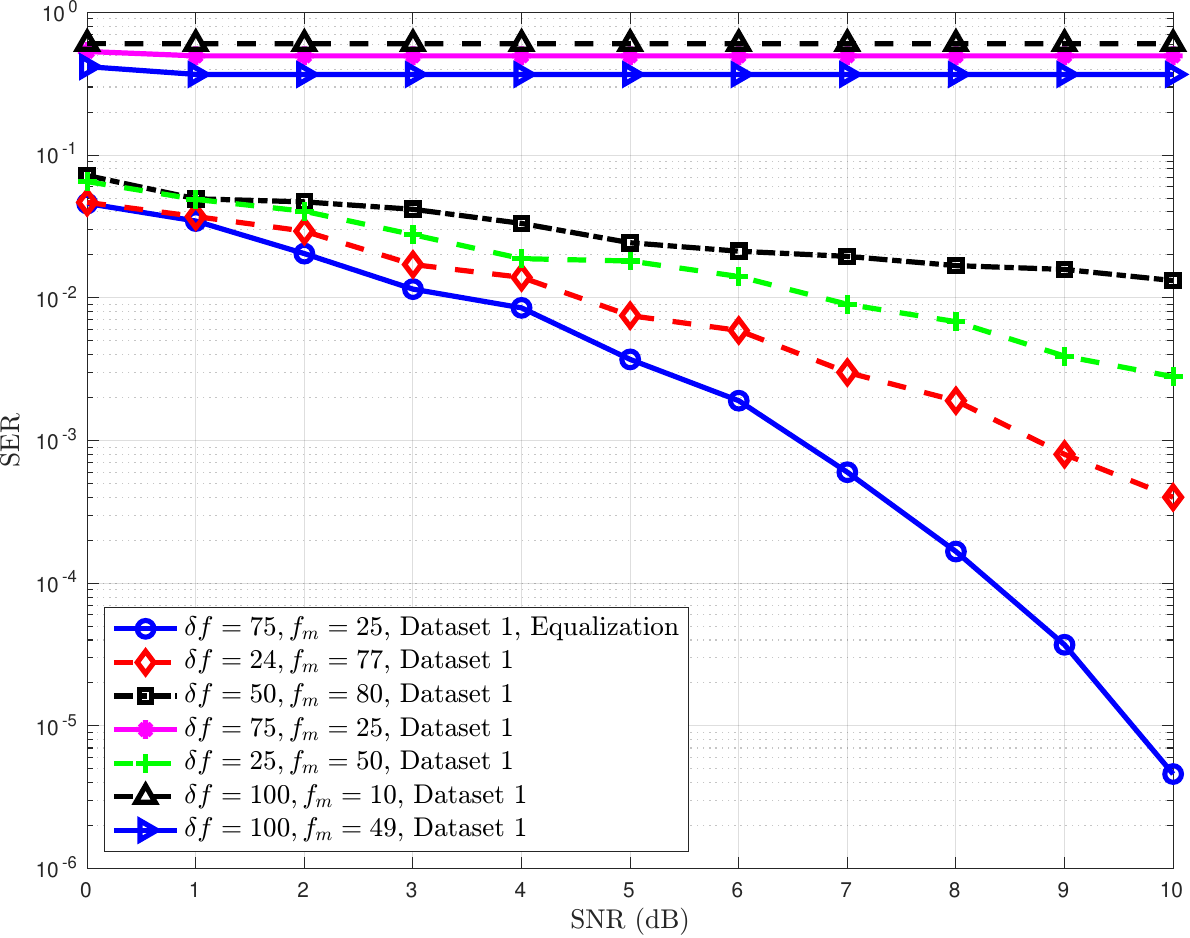}
	\caption{SER performance for QPSK and different parameters in the obfuscating signal. SER performance for Dataset 2 is almost identical.}
	\label{fig:ser-vs-snr_spoofing}
\end{figure}	
	
\textbf{Classification Accuracy in AWGN:} The classification accuracy is presented in Fig.~\ref{fig:accuracy-vs-snr_spoofing}. Without equalization, the results show that for Dataset 1 when $\delta f > f_m$ DNN accuracy is reduced to less than 10\%, but for $\delta f < f_m$ it tops around 38\%. However, for OFDM in Dataset 2 the classification performance keeps reducing when both parameters are increased regardless of their in-between relationship. Still $\delta f$ seems to be the dominant factor even in the OFDM Dataset 2. Recall that $\delta f$ determines the extend of signal spread in the frequency domain, affecting thus more the spectrogram for a given symbol rate. These results provide an idea regarding the settings that one could use so that it can achieve the desired result. We noticed that with our waveform manipulation the classifier confuses almost every modulation type with PAM4 because the obfuscating signal makes it look like this particular modulation type. As expected, the CNN confuses 16-QAM and 64-QAM frames since they share similarities, i.e. 16-QAM is a subset of 64-QAM. For the analog DSB-AM and SSB-AM signals, it is clear from the spectrograms that they have not been altered significantly (since they are amplitude modulations), so the classifier successfully differentiates them. We must also note the performance at the LRx when we use equalization to remove the impact of obfuscation (results only for the worst case settings for $\delta f,f_m$ are illustrated). The results show very good classification performance that is increased as SNR is increased as expected. The reason is that our waveform manipulation method offers a simple way for the LRx to re-create the perturbation and remove it.
	
\textbf{Symbol Error Rate (SER) in AWGN:} Recall that we are interested in the classification performance of the eavesdropper and the demodulation performance of the nominal communication receiver denoted as LRx. To evaluate the impact of obfuscation on the communication performance we focus on digital modulations only. Due to limited space we present the SER in Fig.~\ref{fig:ser-vs-snr_spoofing} for QPSK and Dataset 1. Similar results are obtained for all QAM variants (lower SER of course for higher order modulations) and Dataset 2. When the LRx performs equalization of the obfuscating signal $x_\text{sp}(t)$, there is no performance loss. This is the result of designing the obfuscating signal to generate only phase variations and not amplitude variations not affecting thus the SNR of complex waveforms. When the LRx does not employ any equalization of the obfuscating signal, then we have worse SER. Even though this is undesirable there might be scenarios where we might need to achieve a specific SER and consequently small values for the obfuscating parameters could be tolerated without the need to inform the LRx.
	
\begin{figure}[t]
	\centering
	\includegraphics[width=0.95\linewidth]{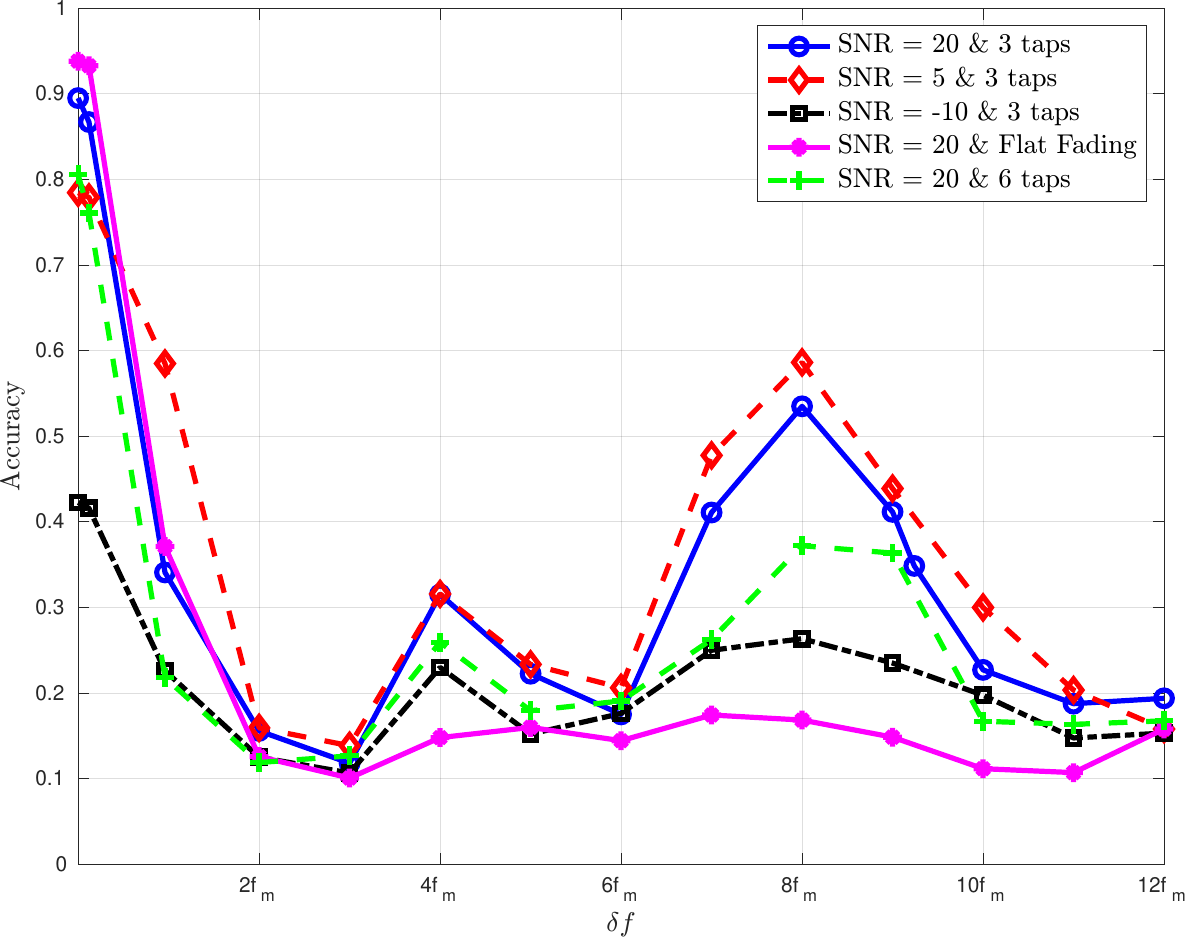}
	\caption{Classification accuracy in frequency selective channels for all modulations in Dataset 1 versus different parameters in the obfuscating waveform.}
	\label{fig:accuracy-fading_spoofing}
\end{figure}

\textbf{Fading Channels:} Results in Fig.~\ref{fig:accuracy-fading_spoofing} concern a frequency selective channel for Dataset 1. This was modeled with different number of taps. Since this dataset consisted of single-carrier communication, maximum likelihood sequence detection (MLSD) equalization was used to combat frequency-selective channel fading in digital communication. Overall we notice the same performance trend for the same SNR and regardless of the channel effect. This means the dominant criterio for classification performance is the use of waveform manipulation or not. However, frequency selectivity seems to add an additional level of distortion that allows slightly better accuracy when compared to flat fading. Specific pairs of $\delta f$ and $f_m$ are more favorable for enabling obfuscation and in this case a $\delta f = 3f_m$ allows the lowest accuracy at the eavesdropper. In the previous figures we can see that the same pair of parameters offers the lowest accuracy for AWGN channels too. Note that SER performance exhibits the same behavior with AWGN channels, that is the parameters that decrease classification accuracy also decrease SER unless we equalize the obfuscating signal.

\textbf{DNN with Adversarial Training:} Due to limited space we present initial results for an adversarially-trained version of the DNN in an AWGN channel in Table~\ref{table:vardescription3}. We present the DNN accuracy versus SNR at the eavesdropper for two different parameter settings of the obfuscating signal at the Tx ($\delta f$ and $f_m$). The result is that adversarial training at the eavesdropper does not improve accuracy significantly for the simple reason that our waveform manipulation strategy for specific $\delta f$ and $f_m$ distorts significantly the phase and frequency of the emitted signal. This makes phase and frequency modulated waveforms look very similar.

\section{Conclusions}
In this paper we explored the use of a new wireless waveform manipulation method for preventing modulation classification with DNN techniques. The proposed waveform is coupled with its careful planting in the transmitted signal so as to prevent the extraction of clean data for adversarial training of the DNN-based classifier. The proposed obfuscating signal works for both analog, single carrier, and OFDM modulations and drops classification performance to even less than 10 \% in AWGN, flat, and frequency-selective channels with careful selection of its parameters. Similar results are also obtained for an adversarially trained DNN.

\begin{flushleft}
	\begin{table}[t] 
		\caption{Accuracy with Adversarial Training in AWGN.} 
		\label{table:vardescription3}
		\begin{center}
			\begin{tabular}{|l|*{8}{p{7mm}|}}
				\hline \hline				
				SNR &    -20 & -10 & 0  & 10 & 20 \\\hline 
				Tx: $\delta f=75,f_m=25$& 9\% & 10\% & 11\% & 11\% &  11\% \\ \hline 
				Tx: $\delta f=24,f_m=77$ & 29\% & 41\% & 44\% & 44\% &  44\%  \\ \hline 												
			\end{tabular}
		\end{center}
		
	\end{table}
\end{flushleft}

	\bibliographystyle{IEEEtran}
	\bibliography{../../../../tony-bib}
	\end{document}